# ANALISIS
# LAPORAN TUGAS AKHIR MAHASISWA DIPLOMA I DARI SUDUT PANDANG KAIDAH KARYA ILMIAH DAN PENGGUNAAN TEKNOLOGI INFORMASI
### (Pada Program Studi Manajemen Informatika DI)

**Leon Andretti Abdillah dan Emigawaty**
Universitas Bina Darma
Jln. Ahmad Yani No. 12, Plaju, Palembang
Pos-el: leon.abdillah@yahoo.com dan emigawaty@yahoo.com

**Abstracts**: The purposes of this research are: 1) to analyze final report from scientific role, 2) the use of information technology (IT), and 3) to conduct academic athmosphere in research area. This research gives contributions to study program of MI-DI, such as: 1) to know the pattern of student final report from scientific role and the use of IT, 2) give input to study program for next final report scheme, and 3) can be used for next research reference. If we look to the quality of final report, there are several focuses to be prepared on tittle, literature review, methodology, results of report, discussion and also conclusion. But for the use of IT is already good but the varian is decrease.

**Keywords**: Scientific Report, Scientific Role, Information Technology.

## 1. PENDAHULUAN

Pendidikan merupakan suatu kebutuhan primer yang sejak dini hingga dewasa hendaknya dirasakan oleh seluruh masyarakat. Hal ini sesuai dengan amanat UUD Negara Kita, anjuran agama, dan menjadi penentu kemajuan suatu bangsa.

Pendidikan pada level perguruan tinggi terdiri atas strata DI, DII, DIII, S1, S2, S3, hingga *postdoctoral*. Bagi sebagian tamatan SMA (Sekolah Menengah Atas) melanjutkan studi ke jenjang S1 dapat dikatakan sebagai impian utama. Tetapi pada kenyataannya banyak aspek yang menyebabkan pilihan strata jatuh pada level di bawah S1, misalnya D-III, D-II, bahkan D-I.

Universitas Bina Darma (UBD) sebagai salah satu Perguruan Tinggi terkemuka di wilayah sumatera, memberikan wadah bagi lulusan SMA untuk mengenyam pendidikan tinggi yang berkualitas dan terjangkau. Untuk level Diploma I, UBD membuka Program Studi Manajemen Informatika D-I dan Komputerisasi Akuntansi D-I, kedua Program tersebut di bawah naungan Fakultas Ilmu Komputer.

Salah satu teknologi yang sedang berkembang dengan pesat saat ini adalah teknologi informasi/komputer. Kemajuan yang berlangsung cepat, dapat ditinjau baik dari segi perangkat keras (*hardware*), perangkat lunak (*software*), maupun perkembangan kualitas sumber daya manusianya (*brainware*). Hal ini dimungkinkan karena teknologi komputer mampu berkolaborasi dengan banyak bidang ilmu lainnya (Abdillah, 2007:1). TI juga sangat berperan dalam pembuatan Laporan Tugas Akhir, misalnya dengan menggunakan paket aplikasi perkantoran, maupun aplikasi perancangan.

Sama seperti jenjang studi pada semua level strata, Diploma I juga melakukan kegiatan yang dilaporkan dalam bentuk Laporan Akhir. Laporan Akhir tersebut merupakan cerminan dari kapabilitas mahasiswa untuk terjun ke dunia nyata dengan menyandang gelar akademik Ahli Pratama (A.P.).

Suatu laporan yang baik hendaknya disusun dengan memperhatikan kaidah-kaidah ilmiah (sesuai kebutuhannya), kemudian didukung dengan Teknologi Informasi (sesuai dengan naungan Fakultas Ilmu Komputer), serta yang tidak kalah pentingnya adalah konsistensi struktur tulisan.

Dengan mempertimbangkan hal-hal tersebut di atas, serta dalam rangka tetap menjaga kualiatas UBD melalui lulusannya (salah satunya D-I), maka peneliti tertarik untuk melakukan telaah dan kajian pada Program Studi Manajemen Informatika Diploma I, khususnya yang berhubungan Laporan Tugas Akhir Mahasiswa.

Berdasarkan latar belakang yang telah diungkapkan di atas, maka perumusan malahan pada penelitian ini adalah "Bagaimana analisis laporan tugas akhir mahasiswa Diploma I dari sudut pandang kaidah karya ilmiah, dan penggunaan teknologi informasi pada Program Studi Manajemen Informatika D-I?"

Agar penelitian ini dapat dilakukan dengan seksama dan terarah serta meminimalkan bias yang mungkin terjadi, maka perlu dilakukan pembatasan khususnya pada Program Studi Manajemen Informatika D-I. Laporan yang akan diteliti dimulai dari angkatan 2005 s.d. 2007.

Penelitian ini dilakukan tidak hanya dalam rangka pelaksanaan penelitian rutin per-semester dilingkungan Fakultas Ilmu Komputer, tetapi juga untuk menjaga *academic athmosphere* agar tetap kondusif. Tujuan penelitian adalah: 1) Menganalisis laporan akhir mahasiswa Diploma I dari sudut pandang kaidah karya ilmiah, penggunaan teknologi informasi, dan konsistensi struktur laporan pada Program Studi Manajemen Informatika D-I, 2) Melakukan penelitian pada Program

Studi Manajemen Informatika D-I, dan 3) Menghidupkan *academic athmosphere* khususnya yang berhubungan dengan bidang penelitian.

Dengan adanya penelitian ini akan memberikan kontribusi kepada Program Studi Manajemen Informatika D-I Fakultas Ilmu Komputer Universitas Bina Darma, yaitu: 1) Mengetahui pola laporan tugas akhir mahasiswa Program Studi Manajemen Informatika Diploma I dari sudut pandang kaidah karya ilmiah, penggunaan teknologi informasi, dan konsistensi struktur laporan, 2) Memberikan masukan pada Program Studi Manajemen Informatika DI yang berhubungan dengan pola laporan tugas akhir mahasiswa, khususnya untuk penerapan pada masa yang akan datang, dan 3) Menambah acuan untuk kajian-kajian lanjutan.

## 2.   TINJAUAN PUSTAKA

Untuk memudahkan proses penelitian yang berhubungan dengan karya ilmiah dilihat dari kaidah ilmiah dan penggunaan teknologi informasi, serta dalam rangka menyamakan persepsi bagai peneliti dan pembaca. Maka ada sejumlah istilah atau terminologi yang penulis coba terangkan sesuai dengan rujukan yang penulis baca/pahami sebagai rujukan.

### 2.1   Analisis

Analisis merupakan penguraian suatu pokok atas berbagai bagiannya dan penelaahan bagianitu sendiri, serta hubungan antar bagian untuk memperoleh pengertian yang tepat dan pemahaman arti keseluruhan (Darminto & Julianty, 2002). Senada dengan hal tersebut, Soemarno (2004) mengungkapkan bahwa Analisis adalah uraian atau usaha mengetahui arti sutau keadaan.

### 2.2   Tugas Akhir

Laporan Tugas Akhir merupakan karya ilmiah yang dibuat oleh mahasiswa pada program Studi Manajemen Informatika D-I sebagai persyaratan untuk mendapatkan gelar akademik A.P. (Ahli Pratama).

Tugas Akhir adalah karya ilmiah yang disusun oleh mahasiswa (;Program Diploma I) berdasarkan hasil penelitian dari analisis data primer dan atau data sekunder sebagai salah satu persyaratan kelulusan program pada program studinya (Politeknik Cahaya Surya).

Berdasarkan pengetahuan di atas, maka peneliti meyimpulkan bahwa laporan Tugas Akhir Mahasiswa Diploma I adalah karya ilmiah yang disusun oleh mahasiswa (;Diploma I) berdasarkan data (primer/sekunder) disusun secara komprehensif (dengan aturan tertentu) untuk persyaratan kelulusan pada programnya.

**2.3    Karya Tulis**

Karya Tulis Mahasiswa merupakan tulisan berisi ide kreatif dan orisinil yang disusun secara komprehensif berdasarkan data akurat (terpercaya) dianalisis secara runtut, tajam dan diakhiri dengan kesimpulan serta saran-saran/rekomendasi (Ariatedja, 2008).

Umam (2005) menyatakan bahwa karya tulis ilmiah merupakan tulisan berupa gagasan kreatif yang disusun secara komprehensif berdasarkan data akurat (terpercaya), dianalisis secara runut, tajam yang diakhiri dengan kesimpulan yang relevan.

Berdasarkan pengetahuan di atas, maka peneliti meyimpulkan bahwa karya tulis adalah karya yang disusun oleh mahasiswa (;Diploma I) berdasarkan data (primer/sekunder) disusun secara komprehensif (dengan aturan tertentu).

**2.4    Kaidah Ilmiah**

Purwaka (2008) berpendapat bahwa Kaidah ilmiah di antaranya adalah teknik pengumpulan data dan analisis datanya akurat. Sementara itu Tim penyusun (2008) mengemukanan bahwa kaidah-kaidah ilmiah adalah Mengemukakan pokok-pokok pikiran, menyimpulkan dengan melalui prosedur yang sistematis dengan menggunakan pembuktian ilmiah/meyakinkan.

Untuk penelitian ini lebih memfokuskan pada teknik pengumpulan data karena seluruh laporan tugas akhir mahasiswa Diploma I tidak menggunakan analisi data.

**2.5    Teknologi Informasi**

Teknologi Informasi (TI) merupakan sekumpulan dari peralatan *hardware*, *software*, dan *brainware* untuk mengolah data menjadi informasi bagi beragam kebutuhan/tujuan dalam batas ruang dan waktu (Abdillah, 2007:4).

Untuk tugas akhir mahasiswa Diploma I Program Studi Manajemen Informatika Fakultas Ilmu Komputer Universitas Bina Darma, tema tugas akhir

untuk laporannya adalah aplikasi berbasis web, komputer grafis, dan multimedia (Tim Penyusun, 2008:74).

### 2.5.1 Perangkat Lunak (*Software*)

Pressman (2002:10) menggambarkan perangkat lunak adalah: 1) perintah (program komputer) yang bila dieksekusi memberikan fungsi dan unjuk kerja seperti yang diinginkan, 2) struktur data yang memungkinkan program memanipulasi informasi secara proporsional, dan 3) dokumen yang menggambarkan operasi dan kegunaan program.

Sementara Abdillah (2009:5) menyatakan Perangkat Lunak komputer merupakan perangkat yang secara nyata tidak dapat di-akses oleh panca indera manusia, namun ia ada dan sangat penting peranannya.

### 2.5.2 Perangkat Lunak Aplikasi (*Application Software*)

Application software is any tool that functions and is operated by means of a computer, with the purpose of supporting or improving the software user's work. In other words, it is the subclass of computer software that employs the capabilities of a computer directly and thoroughly to a task that the user wishes to perform. This should be contrasted with system software (infrastructure) or middleware (computer services/processes integrators), which is involved in integrating a computer's various capabilities, but typically does not directly apply them in the performance of tasks that benefit the user. In this context the term application refers to both the application software and its implementation (http://en.wikipedia.org/wiki/Application_software).

## 3. METODOLOGI PENELITIAN

### 3.1 Lokasi dan Waktu Penelitian

Karena penelitian ini bersifat *Library Research* (Penelitian Kepustakaan), yang dilaksanakan dengan menggunakan literatur (kepustakaan) dari penelitian sebelumnya (berupa laporan tugas akhir mahasiswa Diploma I), maka lokasi penelitian ini dilaksanakan pada Perpustakan Universitas Bina Darma, yang berada di Kampus C.

Waktu yang dibutuhkan untuk melakukan menelaah laporan tugas akhir pada penelitian ini adalah 4 (empat) bulan, dimulai dari bulan Nopember 2008 sampai dengan bulan Februari 2009.

### 3.2 Sumber dan Pengumpulan Data

Mengingat jumlah laporan mahasiswa Diploma I yang ada di lingkungan Universitas Bina Darma relatif banyak (2002 s.d. 2007), penulis tidak akan mengambil data dari semua laporan yang ada. Tetapi akan mengambil sebagian dari populasi tersebut dengan Cara Stratifikasi (*Stratified Random Sampling*). Menurut Umar (2003:138) Cara Stratifikasi dilakukan dengan cara mengelompokkan populasi yang dianggap heterogen menurut karakteristik tertentu menjadi subpopulasi, sehingga setiap kelompok akan memiliki anggota sampel yang relatif homogen.

Sedangkan teknik pengumpulan data yang digunakan adalah: Data Sekunder. Yaitu data yang diperoleh secara tidak langsung, misalnya dari buku-buku, literatur, melalui internet, media cetak dan karangan ilmiah yang ada hubungannya dengan penelitian ini.. Peneliti melakukan kajian terhadap literatur yang berhubungan dengan laporan tugas akhir mahasiswa Diploma I.

### 3.3 Populasi dan Sampel Penelitian

Total populasi laporan tugas akhir mahasiswa Diploma I Universitas Bina Darma (2005 s.d. 2007) adalah 43, namun laporan yang ada pada rak hanya berjumlah 28, dengan sebaran: 1) Angkatan tahun 2002 sebanyak 1 Judul, 2) Angkatan tahun 2003 sebanyak 1 Judul, 3) Angkatan tahun 2004 sebanyak 0 Judul, 4) **Angkatan tahun 2005 sebanyak 13 Judul**, 5) **Angkatan tahun 2006 sebanyak 6 Judul**, dan 6) **Angkatan tahun 2007 sebanyak 7 Judul**.

Subpopulasi yang penulis ambil adalah berdasarkan tahun angkatan, dimulai dari angkatan 2005 s.d. 2007, sebagai berikut: 1) Untuk Angkatan 2005 sebanyak 8 Laporan, 2) Untuk Angkatan 2006 sebanyak 6 Laporan, dan 3) Untuk Angkatan 2007 sebanyak 7 Laporan.

### 3.4 Sifat Penelitian

Berdasarkan Tempat Penelitian, maka suatu penelitian dibedakan menjadi 3 (tiga) (Tim Penyusun, 2008), yaitu: 1) Field Research (Penelitian Lapangan), langsung di lapangan, 2) **Library Research** (Penelitian Kepustakaan), dilaksanakan dengan menggunakan literatur (kepustakaan) dari penelitian sebelumnya, dan 3) Laboratory Research (Penelitian Laboratorium), dilaksanakan pada tempat tertentu / lab, biasanya bersifat eksperimen atau percobaan. Penelitian ini bersifat Library Research (Penelitian kepustakaan).

### 3.5 Alat dan Bahan

Agar penelitian ini dapat dikerjakan dengan baik, maka dibutuhkan sejumlah alat untuk membantu aktivitas pengumpulan, analisis, dan rekapitulasi data, serta bahan yang akan diolah.

**Alat** digunakan untuk mengubah input/data yang diambil agar dapat diubah menjadi sesuatu yang lebih baik. Ia bersifat tidak habis pakai. Pada penelitian ini alat yang digunakam, adalah: 1) Komputer Jinjing (Notebook), 2) Printer, 3) Kamera Digital, 4) Alat Tulis (Pena, pensil, Penghapus), 5) Sistem Operasi Windows, 6) Microsoft Word XP, dan 7) Microsoft Excel.

Sedangkan **bahan** merupakan sesuatu yang berupa input/data yang diambil agar dapat diubah menjadi sesuatu yang lebih baik. Ia bersifat habis pakai. Pada penelitian ini baha yang digunakam, adalah: 1) Data Laporan Tugas Akhir, 2) Kertas, dan 3) Tinta .

### 4. PEMBAHASAN

Sebelum melihat hasil dari penelitian ini ada baiknya kita juga melihat deskripsi lokasi penelitian yang berada di Perpustakaan Universitas Bina Darma yang terletask di Kampus C.

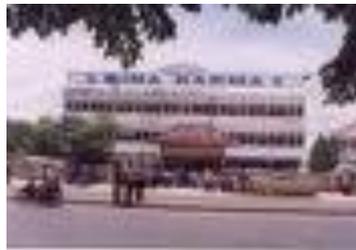

**Gambar 4.1 Kampus C Universitas Bina Darma**

Perpustakaan Universitas Bina Darma merupakan salah satu unit pelayanan teknis yang didanai oleh Bank Pembangunan Asia (ADB). Adapun beberapa bentuk bantuan tersebut diantaranya berupa 465 kopi judul, serta beberapa peralatan penunjang kegiatan perpustakaan.

Jika dilihat dari infrastruktur yang dimiliki oleh perpustakaan Universitas Bina Darma sangatlah memadai, sudah dilengkapi dengan e-library, komputer untuk searching data, sistem pencatatan pengunjung, dll.

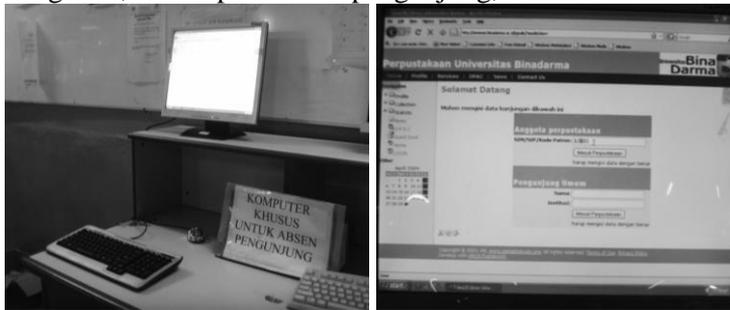

**Gambar 4.2.a Sistem Komputer Pengunjung**

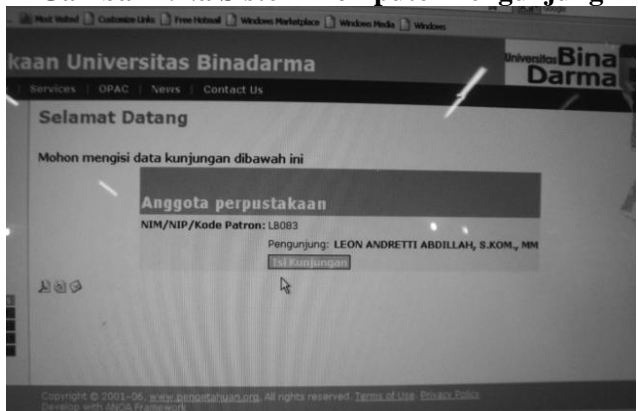

**Gambar 4.2.b Sistem Komputer Pengunjung**

Jenis layanan yang diberikan perpustakaan Universitas Bina Darma antara lain memberikan informasi umum mengenai fasilitas, layanan dan peraturan perpustakaan, membantu pengguna dalam mencari koleksi perpustakaan, memberikan layanan penelusuran artikel atau informasi dari koleksi perpustakaan dan instansi lain, memberikan bimbingan dan pendidikan pemakai, pengenalan fasilitas perpustakaan, layanan perpustakaan, penyediaan fasilitas pendukung serta tata tertib yang ada di perpustakaan. Hal ini dilakukan dengan sebaik mungkin dengan tujuan untuk memberikan kepuasan khususnya bagi para pengguna perpustakaan (Nasir, Antoni, Kurniawan, 2007).

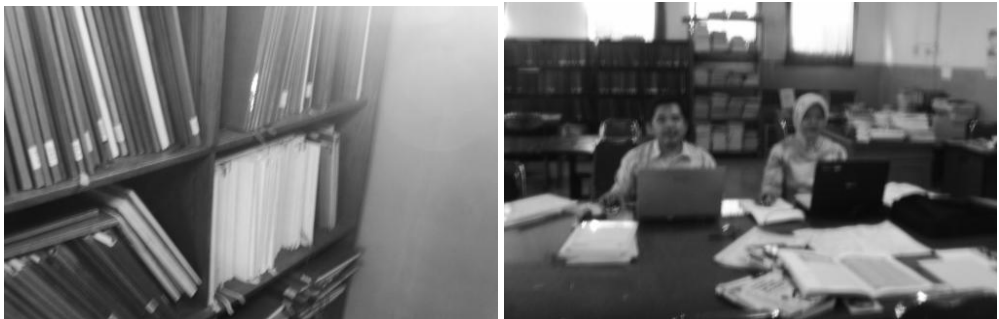

**Gambar 4.3 Rak Laporan Tugas Akhir & Meja Baca**

Laporan Tugas Akhir Mahasiswa merupakan salah satu koleksi dari Perpustakaan Universitas Bina Darma. Untuk Program Diploma I, laporan Tugas Akhir Mahasiswa menggunakan warna putih.

## 4.1 Hasil

Setelah melakukan telaah terhadap laporan yang ada pada perpustakaan, dihasilkan sejumlah temuan yang disesuaikan dengan tinjauan yang penulis gunakan dan tertera di Bab II.

## 4.2 Rekapitulasi Laporan Tugas Akhir MIDI

Berdasarkan jumlah laporan yang ada di rak perpustakaan Universitas Bina Darma yang berada di Kampus C, terdapat 28 laporan tugas akhir mahasiswa Diploma I, yang terdiri atas: 1) Angkatan tahun 2002 sebanyak 1 Judul, 2) Angkatan tahun 2003 sebanyak 1 Judul, 3) Angkatan tahun 2004 sebanyak 0 Judul, 4) Angkatan tahun 2005 sebanyak 13 Judul, 5) Angkatan tahun 2006 sebanyak 6 Judul, dan 6) Angkatan tahun 2007 sebanyak 7 Judul.

### 4.2.1 Analisis Struktur Laporan Tugas Akhir MIDI

Struktur secara umum laporan-laporan yang ada sudah dapat dikatakan baik, karena semua laporan sudah memiliki pola yang sama untuk struktur laporannya, yang berbeda hanyalah keberadaan bagian tata kerja (Bab III) sebagian dimasukkan ke dalam Bab I.

```
COVER
HALAMAN JUDUL
```

```
HALAMAN PENGESAHAN
HALAM MOTTON DAN PERSEMBAHAN
KATA PENGANTAR
DAFTAR ISI
DAFTAR GAMBAR
DAFTAR TABEL
BAB I PENDAHULUAN
BAB II TINAJUAN PUSTAKA
BAB III HASIL DAN PEMBAHASAN
BAB IV KESIMPULAN DAN SARAN
DAFTAR PUSTAKA
LAMPIRAN
```

**Tabel 4.1 Rekapitulasi Jumlah Halaman Laporan Tugas Akhir**

| Tahun | LTA 1 | LTA 2 | LTA 3 | LTA 4 | LTA 5 | LTA 6 | LTA 7 | LTA 8 |
|---|---|---|---|---|---|---|---|---|
| 2005 | x + 26 | ix + 18 | ix + 14 | viii + 20 | vii + 20 | vi + 24 | vii + 26 | vii + 21 |
| 2006 | vi + 24 | vi + 31 | v + 35 | vi + 11 | v + 8 | vi + 36 | | |
| 2007 | ix + 39 | viii + 43 | viii + 30 | ix + 37 | ix + 41 | ix + 26 | ix + 32 | |

(Sumber: hasil telaah data sekunder)

Berdasarkan tabel 4.1 di atas, maka dapat diambil rentang jumlah halaman dari masing-masing tahun angkatan untuk halaman muka ditambah dengan halaman isi, adalah sebagai berikut: 1) Angkatan 2005 (vi - ix) + (14 - 26), 2) Angkatan 2006 (v - vi) + (8 - 36), dan 3) Angkatan 2007 (viii - ix) + (26 - 43).

**4.2.2 Analisis Judul Laporan Tugas Akhir MIDI**

Diketahui bahwa suatu judul yang baik cenderung singkat, biasanya terdiri atas 8 – 12 kata. Dari seluruh sampel laporan, maka didapatkan rekapitulasi jumlah judul kata pada judul (Tabel 4.2).

**Tabel 4.2 Rekapitulasi Jumlah Kata Judul Laporan Tugas Akhir**

| Tahun | LTA 1 | LTA 2 | LTA 3 | LTA 4 | LTA 5 | LTA 6 | LTA 7 | LTA 8 | Rata-rata |
|---|---|---|---|---|---|---|---|---|---|
| 2005 | 7 | 7 | 10 | 6 | 5 | 4 | 4 | 6 | 6.13 |
| 2006 | 7 | 9 | 9 | 9 | 8 | 9 | - | - | 8.50 |
| 2007 | 6 | 5 | 10 | 7 | 5 | 5 | 8 | - | 6.57 |

(Sumber: hasil telaah data sekunder)

Dari ketiga angkatan, secara rata-rata hanya angkatan 2006 saja yang sudah baik dari jumlah kata judul. Untuk angkatan 2005 dan 2007 jumlah kata judulnya secara rata-rata masih di bawah standar.

### 4.2.3  Analisis Pendahuluan Laporan Tugas Akhir MIDI

Secara umum dari ketiga angkatan, untuk bagian Bab I Pendahuluan sudah baik, karena dari semua laporan tugas akhir memiliki pendahuluan, dengan struktur isi:

```
BAB I PENDAHULUAN
   Latar Belakang
   Rumusan Masalah
   Batasan Masalah
   Tujuan dan Manfaat
   Prosedur Pembuatan
```

Hanya saja untuk subbagian Prosedur Pembuatan, akan lebih baik jika diletakkan di bagian Bab III Tata Kerja.

**Tabel 4.3 Rekapitulasi Pendahuluan Laporan Tugas Akhir**

| Tahun | LTA 1 | LTA 2 | LTA 3 | LTA 4 | LTA 5 | LTA 6 | LTA 7 | LTA 8 | Rata-rata |
|---|---|---|---|---|---|---|---|---|---|
| 2005 | 1 | 1 | 1 | 1 | 1 | 1 | 1 | 1 | 1 |
| 2006 | 1 | 1 | 1 | 1 | 1 | 1 | - | - | 1 |
| 2007 | 1 | 1 | 1 | 1 | 1 | 1 | 1 | - | 1 |

(Sumber: hasil pencatatan terhadap laporan tugas akhir)
Catatan : 1 = Ada; 0 = Tidak Ada

Berdasarkan rekapitulasi tabel 4.3 dapat dilihat bahwa secara umum semua bagian pendahuluan laporan tugas akhir sudah lengkap.

### 4.2.4  Analisis Teori Laporan Tugas Akhir MIDI

Untuk bagian Bab II Teori (Tinjuan Pustaka) Penulis tidak melihat apakah teori yang digunakan itu tepat atau tidak, namun lebih kepada ada atau tidaknya sumber rujukan yang digunakan.

**Tabel 4.4 Rekapitulasi Teori Laporan Tugas Akhir**

| 2005 | 1 | 1 | 0 | 0 | 1 | 1 | 0 | 0 | 0.5 |
|------|---|---|---|---|---|---|---|---|-----|
| 2006 | 1 | 0 | 1 | 0 | 1 | 0 | - | - | 0.5 |
| 2007 | 1 | 1 | 1 | 1 | 1 | 1 | 1 | - | 1   |

(Sumber: hasil telaah data sekunder)
Catatan : 1 = Ada Sumber; 0 = Tidak Ada Sumber

Dari hasil rekapitulasi yang tertera pada Tabel 4.4 diketahui bahwa pada Angkatan 2005 dan 2006 hanya 50% saja laporan yang mencantumkan sumbernya. Sedangkan pada Angkatan 2007 sudah 100% laporan yang mencantumkan sumbernya.

Namun angka ini hanya menampilkan keberadaan sumbernya saja, mengenai kecocokan sumber tersebut dengan yang tertera pada Daftar Pustaka akan ditampilkan pada bagian 4.1.9.

### 4.2.5 Analisis Tata Kerja Laporan Tugas Akhir MIDI

Pada bagian ini Penulis membuat daftar yang sebaiknya ada pada bagian Tata kerja suatu laporan ilmiah, yaitu: 1) Objek Penelitian, 2) Bahan, 3) Peralatan, 4) Cara, dan 5) Rancangan/Metode Penelitian/Kegiatan.

**Tabel 4.5 Rekapitulasi Tata Kerja Laporan Tugas Akhir**

| Tahun | LTA 1 | LTA 2 | LTA 3 | LTA 4 | LTA 5 | LTA 6 | LTA 7 | LTA 8 | Rata-rata |
|-------|-------|-------|-------|-------|-------|-------|-------|-------|-----------|
| 2005  | 1     | 1     | 1     | 1     | 1     | 1     | 1     | 1     | 1         |
| 2006  | 1     | 1     | 1     | 1     | 1     | 1     | -     | -     | 1         |
| 2007  | 1     | 1     | 1     | 1     | 1     | 1     | 1     | -     | 1         |

(Sumber: hasil telaah data sekunder)
Catatan : 0 = Tidak Ada, 1 = Ada, 2 = Lengkap

Dari hasil rekapitulasi yang tertera pada Tabel 4.5 diketahui bahwa pada seluruh 100% sudah ada unsur Tata Kerja, namun belum lengkap atau masih ada bagian yang belum dicantumkan.

### 4.2.6 Analisis Hasil dan Pembahasan Laporan Tugas Akhir MIDI

Pada bagian ini Penulis membuat daftar yang sebaiknya ada pada bagian Hasil suatu laporan ilmiah, yaitu: 1) Ada hasil (berupa deskripsi), dan 2) Ada hasil (berupa deskripsi + gambar/tabel).

**Tabel 4.6.a Rekapitulasi Hasil Laporan Tugas Akhir**

| Tahun | LTA 1 | LTA 2 | LTA 3 | LTA 4 | LTA 5 | LTA 6 | LTA 7 | LTA 8 | Rata-rata |
|---|---|---|---|---|---|---|---|---|---|
| 2005 | 1 | 2 | 1 | 1 | 1 | 1 | 1 | 1 | 1.13 |
| 2006 | 2 | 2 | 2 | 2 | 1 | 1 | - | - | 1.67 |
| 2007 | 1 | 1 | 1 | 1 | 1 | 1 | - |   | 1.00 |

(Sumber: hasil telaah data sekunder)
Catatan : 0 = Tidak Ada, 1 = Ada, 2 = Lengkap

Dari hasil rekapitulasi yang tertera pada Tabel 4.6.a diketahui bahwa: Untuk Angkatan 2005 seluruh 100% sudah ada hasil, namun hanya 13% laporan yang menampilkan hasil (berupa gambar/tabel) pada subbagian hasil. Untuk Angkatan 2006 seluruh 100% sudah ada hasil, dan sudah 67% laporan yang menampilkan hasil (berupa gambar/tabel) pada subbagian hasil. Untuk Angkatan 2007 seluruh 100% sudah ada hasil, namun hanya 0% laporan yang menampilkan hasil (berupa gambar/tabel) pada subbagian hasil.

Pada bagian ini Penulis membuat daftar yang sebaiknya ada pada bagian Pembahasan suatu laporan ilmiah, yaitu: 1) Ada pembahasan (berupa deskripsi), dan 2) Ada pembahasan (berupa deskripsi + gambar/tabel) sesuai dengan data yang diperoleh (Tabel atau Gambar).

**Tabel 4.6.b Rekapitulasi Pembahasan Laporan Tugas Akhir**

| Tahun | LTA 1 | LTA 2 | LTA 3 | LTA 4 | LTA 5 | LTA 6 | LTA 7 | LTA 8 | Rata-rata |
|---|---|---|---|---|---|---|---|---|---|
| 2005 | 1 | 1 | 1 | 1 | 2 | 1 | 2 | 1 | 1.25 |
| 2006 | 1 | 1 | 1 | 1 | 1 | 1 | - | - | 1.00 |
| 2007 | 1 | 1 | 1 | 1 | 1 | 1 | 1 | - | 1.00 |

(Sumber: hasil telaah data sekunder)
Catatan : 0 = Tidak Ada, 1 = Ada, 2 = Lengkap

Dari hasil rekapitulasi yang tertera pada Tabel 4.6.b diketahui bahwa: Untuk Angkatan 2005 seluruh 100% sudah ada pembahasan, namun hanya 25% laporan yang pembahasannya sesuai (gambar/tabel) subbagian hasil. Untuk Angkatan 2006 dan Angkatan 2007 seluruh 100% sudah ada hasil, namun pembahasannya belum sesuai dengan (gambar/tabel) subbagian hasil.

### 4.2.7 Analisis Kesimpulan dan Saran Laporan Tugas Akhir MIDI

Pada bagian ini Penulis membuat daftar yang sebaiknya ada pada bagian Hasil suatu laporan ilmiah, yaitu: 1) Ada kesimpulan (kurang sesuai tema), dan 2) Ada kesimpulan (sesuai tema).

**Tabel 4.7 Rekapitulasi Kesimpulan Laporan Tugas Akhir**

| Tahun | LTA 1 | LTA 2 | LTA 3 | LTA 4 | LTA 5 | LTA 6 | LTA 7 | LTA 8 | Rata-rata |
|---|---|---|---|---|---|---|---|---|---|
| 2005 | 1 | 1 | 1 | 1 | 2 | 2 | 1 | 1 | 1.25 |
| 2006 | 1 | 1 | 1 | 1 | 1 | 2 | - | - | 1.17 |
| 2007 | 1 | 1 | 1 | 1 | 1 | 1 | 1 | - | 1.00 |

(Sumber: hasil telaah data sekunder)
Catatan : 0 = Tidak Ada, 1 = Ada, 2 = Lengkap

Dari hasil rekapitulasi yang tertera pada Tabel 4.7 diketahui bahwa: Untuk Angkatan 2005 seluruh 100% sudah ada kesimpulan, namun hanya 25% laporan yang kesimpulannya sesuai tema. Untuk Angkatan 2006 seluruh 100% sudah ada kesimpulan, namun hanya 17% laporan yang kesimpulannya sesuai tema. Angkatan 2007 seluruh 100% sudah ada hasil, namun hanya sebesar 0% kesimpulannya yang sesuai dengan tema.

### 4.2.8 Analisis Daftar Pustaka Laporan Tugas Akhir MIDI

Pada bagian ini Penulis membuat daftar yang sebaiknya ada pada bagian "Daftar Pustaka" suatu laporan ilmiah, kriteria ada tiga: 1) Tidak Ada rujukan (tidak terdaftar pada laporan), 2) Ada rujukan (tidak terdaftar pada laporan), dan 3) Ada rujukan (terdaftar pada laporan).

**Tabel 4.8 Rekapitulasi Daftar Pustaka Laporan Tugas Akhir**

| Tahun | LTA 1 | LTA 2 | LTA 3 | LTA 4 | LTA 5 | LTA 6 | LTA 7 | LTA 8 | Rata-rata |
|---|---|---|---|---|---|---|---|---|---|
| 2005 | 1 | 2 | 1 | 1 | 1 | 1 | 1 | 1 | 1.13 |
| 2006 | 1 | 1 | 1 | 1 | 2 | 1 | - | - | 1.17 |
| 2007 | 1 | 1 | 1 | 1 | 1 | 1 | 1 | - | 1.00 |

(Sumber: hasil telaah data sekunder)
Catatan : 0 = Tidak Ada, 1 = Ada, 2 = Lengkap (pada naskah dan daftar pustaka ada)

Dari hasil rekapitulasi yang tertera pada Tabel 4.8 diketahui bahwa: Untuk Angkatan 2005 seluruh 100% sudah ada daftar pustaka, namun hanya 13% laporan yang daftar pustakanya sesuai dengan yang tertera pasa laporan. Untuk Angkatan 2006 seluruh 100% sudah ada daftar pustaka, namun hanya 17% laporan yang daftar pustakanya sesuai dengan yang tertera pada laporan. Angkatan 2007 seluruh 100% sudah ada daftar pustaka, namun hanya sebesar 0% daftar pustakanya yang sesuai dengan yang tertera pada laporan.

**4.2.9 Analisis Penggunaan Teknologi Informasi Laporan Tugas Akhir MIDI**

Teknologi Informasi yang diperhatikan oleh Penulis adalah yang digunakan oleh mahasiswa untuk mengerjakan topik laporannya. Secara spesifik teknologi informasi yang ditelaah adalah pada aspek *software*. Secara umum hasial rekapitulasi penggunaan *software* untuk menyelesaikan tema laporan dapat dilihat pada tabel 4.9.

**Tabel 4.9 Rekapitulasi Penggunaan Teknologi Informasi**

| Angkatan | *Software* yang digunakan |
|---|---|
| 2005 | Macromedia Dreamweaver MX |
|  | Adobe Photoshop |
|  | Microsoft Frontpage |
|  | Swish |
| 2006 | Adobe PhotoShop 7 |
|  | Corel Draw |
| 2007 | Macromedia Flash MX 2004 |

(Sumber: hasil telaah data sekunder)

Berdasarkan rekapitulasi yang tertera pada Tabel 4.9 penggunaan *software* untuk membuat aplikasi yang sesuai dengan tema tugas akhirnya cukup beragam (sesuai dengan Angkatannya). Namun perlu diperhatikan juga untuk meng-*update* versi dari *software* yang digunakan.

Tema tugas akhir juga hendaknya diperkaya, bahkan jika perlu mahasiswa dilibatkan seperti kegiatan magang atau praktek kerja lapangan pada perusahaan yang sesungguhnya.

**4.3 Analisis Pembahasan**

Untuk penggunaan kaidah ilmiah yang masih belum sepenuhnya sesuai hal ini disebabkan karena mahasiswa memang tidak mendapatkan mata kuliah metode penelitian/penulisan ilmiah.

Jumlah Dosen Pembimbing perlu ditingkatkan jumlahnya dengan memperhatikan keahlian serta minatnya. Dosen Pembimbing perlu meningkatkan peranannya dalam mengarahkan aturan penulisan ilmiah dalam kegiatan konsultasi.

Ketua Program Studi dapat menyiapkan menyiapkan *template* bagi mahasiswa, serta meningkatkan pemeriksaan/pengontrolan kualitas Laporan Tugas Akhir Mahasiswa.

Sedangkan untuk penggunaan Teknologi Informasi khususnya *software* untuk menunjang aplikasi tugasnya secara umum sudah baik, tetapi perlu juga diperhatikan versi dari *software* tersebut hendaknya menggunakan versi yang terbaru.

Perlu adanya pengelompokan (berdasarkan jumlah mahasiswa) yang akan menentukan jumlah/varian *software* yang digunakan pada Tugas Akhir, sehingga tema serta jenis *software* yang digunakan juga akan beragam.

## 5. SIMPULAN

Setelah mencermati temuan yang penulis rekap dan telah disajikan pada Bab IV, maka penulis dapat mengambil sejumlah simpulan yang berhubungan dengan Laporan Tugas Akhir Mahasiswa Diploma I jika ditinjau dari Kaidah-kaidah Ilmiah dan Penggunaan Teknologi Informasi. Simpulannya adalah sebagai berikut:
1) Ada sejumlah hal yang perlu diperhatikan dalam memperbaiki kualitas Laporan Tugas Akhir Mahasiswa Diploma I, yaitu:
    a. Judul laporan masih terlalu singkat
    b. Sumber pada laporan disesuaikan dengan daftar pustaka
    c. Bab III hendaknya berisi tentang Tata Kerja/Metodologi Penelitian.
    d. Hasil dari laporan selain narasi juga disajikan dalam bentuk gambar atau tabel.
    e. Pembahasan dibuat dengan mengacu kepada tabel atau gambar yang dihasilkan.
    f. Kesimpulan dibuat dengan memperhatikan kesesuaian dengan hasil dan pembahasan yang mengacu kepada tema.
2) Penggunaan Teknologi Informasi sudah baik dan sejalan dengan tema yang diselenggarakan oleh program studi pertahun. Walaupun pada angkatan-angkatan terakhir (2005, 2006, 2007) cenderung menurun varian-nya.
3) Sedangkan hal-hal yang perlu diperbaiki unutk ke depan, adalah sebagai berikut:

a. Dosen pembimbing dan Ketua Progam Studi hendaknya memberikan perhatian yang lebih teliti terhadap hasil laporan tugas akhir mahasiswa Manajemen Informatika Diploma I, serta adanya penyebaran pembimbing.
b. Jika diperlukan dapat dibuatkan *template* bagi mahasiswa untuk memandu pembuatan laporan tugas akhirnya.
c. Tema pertahun hendaknya dibuat dengan sebaran yang baik, sehingga menambah warna bagi laporan tugas akhir mahasiswa.
d. Jika perlu laporan mahasiswa disinkronkan dengan sistem sertifikasi yang sesuai dengan jenjang Diploma I, sehingga mahasiswa akan tamat dengan mendapatkan Ijazah A.P. beserta sertifikat keahlian di bidang komputer.
e. *Software* yang digunakan hendaknya sudah menggunakan edisi terbaru sehingga mahasiswa memiliki pengetahuan yang *up to date* terhadap perkembangan teknologi *software* komputer.

# DAFTAR RUJUKAN